\documentclass[aps,twocolumn,prd]{revtex4}
\tighten

\input epsf
\usepackage{graphicx}
\newcommand{\beq}{\begin{equation}}
\newcommand{\beqa}{\begin{eqnarray}}
		  \newcommand{\eeq}{\end{equation}}
\newcommand{\eeqa}{\end{eqnarray}}

\newcommand{\lsim}{\lesssim}
\newcommand{\gsim}{\gtrsim}

\newcommand{\vect}[1]{\mbox{\boldmath${#1}$}}
\newcommand{\lmk}{\left(}
\newcommand{\rmk}{\right)}

\newcommand{\lkk}{\left[} 
\newcommand{\rkk}{\right]}

\newcommand{\vej}{{\vect j}}

\begin{document}

\title{Highly Eccentric Kozai Mechanism and GW Observation for Neutron Star Binaries  } 
%
%
\author{Naoki Seto}
\affiliation{Department of Physics, Kyoto University
Kyoto 606-8502, Japan
}
\date{\today}

%
%
%
%
\begin{abstract}
The Kozai mechanism for a hierarchical triple system could reduce the merger time 
of inner eccentric binary emitting gravitational waves (GWs), and has been 
qualitatively explained with the  secular theory that is derived by averaging 
short-term orbital revolutions. However, with the secular theory,  the minimum 
value of the inner pericenter distance  could be  excessively
limited by the averaging operation.    Compared with 
traditional predictions, the actual 
evolution of an eccentric inner binary could be   accompanied by (i) a higher 
characteristic frequency of the pulse-like GWs around its pericenter passages,  
and (ii) a larger residual 
eccentricity at its final inspiral phase. These findings would be important for 
 GW astronomy with the forthcoming advanced detectors.

\end{abstract}
\maketitle

\section{introduction}

Today, large-scale laser interferometers  are under development to attain a world-wide network of second-generation GW detectors \cite{adv}. Their overall sensitivities will be
improved  by a factor of $\sim  10$,  with drastic noise reduction at the lower 
frequency regime down to $\sim 10$Hz \cite{adv}. Accordingly, understandings of basic properties
of potential astrophysical sources
 have become significant, more than ever.

One of the most promising targets of these detectors is inspiral of a 
neutron star binary (NSB), and in this paper we focus our attention to GW 
observation for NSBs.  
From identified samples in our Galaxy, NSBs are 
expected to have  very small 
residual eccentricities ($O(10^{-5})$) around 10Hz \cite{Lorimer:2008se,Krolak:1995md}.

Meanwhile, it has been pointed out that the Kozai mechanism might play  
important roles for compact binary mergers \cite{Blaes:2002cs,Miller:2002pg,Wen:2002km}. This mechanism works for 
hierarchical triple systems, and oscillates  pericenter distances of inner binaries, 
due to exchange of angular momenta between the inner/outer orbits \cite{Kozai:1962zz}. 
This characteristic feature  can be qualitatively understood with the 
secular theory for which, following a perturbative method in  analytical mechanics,  we effectively average out short-term fluctuations 
associated with 
both 
the inner/outer orbital revolutions \cite{gs,Ford:1999mi}.  Since energy loss due to GW emission depends strongly 
on pericenter distance, the Kozai mechanism can largely reduce the merger time of an 
inner NSB of  a triple system. This 
interesting possibility   has been actively discussed mostly with the secular 
theory including the averaging operations  
  \cite{Blaes:2002cs,Miller:2002pg,Wen:2002km} (see also \cite{Antonini:2012ad}).

In this paper, we  show
that,   for a highly eccentric inner binary emitting GWs, there is a breakdown of 
the secular theory or orbital-averaged approximation, in comparison to the full 
numerical integration.
To handle evolution of such a  binary,  we need to properly resolve the two orbital revolutions  without taking their averages. 
For an inner NSB, this could results in (i) a higher characteristic frequency of the pulse-like GWs around its
  pericenter passages, (ii) a  higher residual eccentricity at its final inspiral 
  phase, and (iii) a  shorter merger time. All of these changes could be 
  more  
  than one order of magnitude.  Our findings (i) and (ii) are significant for the
  advanced  detectors and  their data analyses.  While quantitative evaluation for the merger rate requires detailed astronomical assumptions 
  and is beyond scope of this paper, the last one (iii) indicates a higher  
  merger rate for NSBs of triples  in star clusters \cite{Miller:2002pg}.  This 
  is because the outer 
  third body would be frequently perturbed there. 

 In this paper, 
 we  only discuss relativistic effects for hierarchical triples, but tidal effects  around planets  also depend strongly on orbital 
distance \cite{ssd} (see also \cite{Katz:2012je} for collisions of white 
 dwarf binaries).  For extra-solar planetary systems  ({\it e.g.} Hot 
Jupiters \cite{holman,naoz}),  an investigation similar to this work 
 would be worth considering.

\section{secular theory}

We study evolution of a hierarchical triple system of point masses $m_0$, $m_1$ 
and $m_2$.  We basically use the geometrical units with $G=c=(m_0+m_1+m_2)=1$.   
The inner binary is composed by $m_0$ and $m_1$, and we denote its semimajor 
axis by $a_1$ and its instantaneous orbital separation by $d_1$. In 
 the next section, we also introduce astrophysical units, considering 
 $m_0$-$m_1$ as a NSB.  For the outer third body $m_2$, we define its  semimajor axis $a_2$, relative to the mass center of the inner binary (total mass $M_1\equiv m_0+m_1$).
Likewise,  we use the labels  $j=1$ and 2 for the
inner and outer orbital elements ({\it e.g.} $e_1$ for the inner eccentricity), and 
assume hierarchical orbital configurations with $\alpha\equiv a_1/a_2\ll 1$.

First, we briefly discuss long-term secular evolution of the  triple 
system in Newtonian dynamics, following the approach developed by von Zeipel \cite{gs}. By 
suitably using canonical transformations, we effectively average the short-term 
fluctuations associated with both the inner and outer mean anomalies $l_1$ and 
$l_2$ (the instantaneous angular positions of the inner/outer 
point masses \cite{ssd}).  
The relevant Hamiltonian after the averaging operations, can be evaluated  perturbatively with the expansion parameter $\alpha\ll 1$.  
The leading order (quadrupole) term $H_{qd}=O(\alpha^2)$ is given by \cite{Blaes:2002cs,Miller:2002pg,Wen:2002km,Ford:1999mi}
\beq
H_{qd}=C_{qd} \lkk (2+3e_1^2)(1-3\theta^2)-15e_1^2(1-\theta^2)\cos2\omega_1   \rkk
\eeq
with
$C_{qd}\equiv {m_0 m_1 m_2 \alpha^2}/{[16 M_1 a_2 (1-e_2^2)^{3/2}]} $ and  
 the argument of the inner  pericenter  $\omega_1$   \cite{ssd}. 
Here we define $\theta\equiv \cos I$ with the opening  angle $I$ 
between the inner/outer orbital angular momentum vectors (identical to the angle 
$i$ in \cite{Ford:1999mi}).  
We denote the next order (octupole) term by $H_{oc}(=O(\alpha^3))$ \cite{Ford:1999mi}. For our 
secular analysis of the inner binary, we keep up to this term for the 
gravitational perturbation externally induced by  $m_2$. But  there 
exists a relation $H_{oc}\propto  (m_0-m_1)$,  resulting in $H_{oc}=0$ for  
$m_0=m_1$ \cite{Ford:1999mi}.  Later we use this property to examine possible effects of the  sub-leading terms.

Next, we mention general relativistic corrections to the system, using the 
post-Newtonian (PN) expansion. The lowest order (1PN) term $H_{1pn}$ for our 
hierarchical configuration is obtained after averaging the 
inner mean anomaly $l_1$ as \cite{Blaes:2002cs,Miller:2002pg,Wen:2002km,gs}
\beq
H_{1pn}=-\frac{3m_0m_1M_1}{a_1^2 (1-e_1^2)^{1/2}}. 
\eeq
At this stage, our 
effective Hamiltonian $H_c$ for the secular evolution is given by
\beq
H_{c}=H_{qd}+H_{oc}+H_{1pn},
\eeq
and the system is conservative (thus putting the subscript "c" above) \cite{Blaes:2002cs,Miller:2002pg,Wen:2002km}.  Using canonical equations and transformations of variables, we have {\it e.g.}
\beq
\lmk \frac{d\omega_1}{dt}\rmk_{c}=6 C_{qd}\lmk \frac{4\theta^2}{G_1}+\cdots \rmk +{\rm O.T.}+\frac{3}{a_1(1-e_1^2)} \lmk \frac{M_1}{a_1}  \rmk^{3/2},\label{om}
\eeq
\beq
\lmk \frac{de_1}{dt}\rmk_{c}=30C_{qd}  
\frac{e_1(1-e_1^2)}{G_1}(1-\theta^2)\sin2\omega_1+{\rm O.T.} ,
\eeq
 $(da_1/dt)_{c}=(da_2/dt)_{c}=dH_c/dt=0$ and the scaling relations 
 $(de_2/dt)_{c}={\rm O.T.}$ and $(d\omega_2/dt)_{c}=O(\alpha^2)$.  Here we defined
$G_1=m_0 m_1 \lkk {a_1(1-e_1^2)}/{M_1}  \rkk^{1/2}$ and put $\rm O.T.$ for terms 
 of $O(\alpha^3)$
originating from $H_{oc}$ \cite{Blaes:2002cs,Wen:2002km}. 
The total angular momentum is conserved with $\frac{d}{dt}\sqrt{a_2(1-e_2^2)}=O(\alpha^3)$  for 
 the magnitude of the 
outer one.

The triple system becomes 
dissipative at the 2.5PN order, due to emission of GWs.  
Given our hierarchical configuration, the dissipation predominantly works for the 
inner binary, and we include its effects only for   $a_1$ and 
$e_1$, using standard formulae for isolated eccentric binaries \cite{Peters:1964zz}. 
  Combining these with the conservative contributions, we can  write down the 
  final expressions for the secular evolution such as $d\omega_1/dt=(d\omega_1/dt)_c$,
\beq
\frac{da_1}{dt}=-\frac{64 m_0 m_1 M_1 }{5 a_1^3 (1-e_1^2)^{7/2}} \lmk 1+\frac{73}{24}e_1^2+\frac{37}{96}e_1^4  \rmk,\label{da}
\eeq
\beq
\frac{de_1}{dt}=-\frac{304 m_0 m_1 M_1 
e_1}{15 a_1^4 (1-e_1^2)^{5/2}} \lmk 1+\frac{121}{304}e_1^2  \rmk+\lmk \frac{de_1}{dt}\rmk_{c},\label{de}
\eeq
which have strong dependencies on $1-e_1$. We also have $a_2=const$.   These secular 
equations have been widely used for analyzing  long-term evolutions of 
relativistic hierarchical triple systems \cite{Blaes:2002cs,Wen:2002km}.

\section{numerical results}
In this section, we numerically discuss the Kozai mechanism for relativistic 
hierarchical triples, first using the secular equations and then directly 
integrating the PN equations for three-body systems.  While a triple system has 
many parameters, we  fix most of them to concisely explain our new findings.

In our geometrical units, we fix the masses at $M_1=0.2$,  $m_2=0.8$, and the initial 
orbital parameters at $a_{1}=3.57\times 10^5$, $a_{2}=60 a_{1}=2.14\times 10^7\equiv a_{2i}$ 
({\it i.e.} initially $\alpha=1/60$), $e_{1}=0.2$ and $e_{2}=0.6$.  We 
also set  the initial angular variables  at
$\omega_1=\pi/2$ and $\Omega_1=\omega_2=0$ ($\Omega_1$: the longitude of the inner 
ascending node \cite{ssd}). 
 For our study, the remaining important parameter is the initial inclination $I_{i}$. 
 We explore the regime $I_{i}\sim 90^\circ$ for which an inner binary can merge 
 in a short time (also preferable  for costly direct calculations).

For actual astrophysical system, we presume that the inner binary is  a NSB with 
their total mass  $M_1=2.8 M_\odot$.  Then the initial axes 
correspond to $a_1=0.05$ AU and $a_2=a_{2,i}\equiv3$AU. 
Below, instead 
of  the direct  time variable  $t$,  we use the effective outer revolution cycles 
$N_{2}\equiv t/P_{2i}$ defined with the initial orbital period $P_{2i}=2\pi 
a_{2i}^{3/2}$ (corresponding to 1.38yr).
The primary GW frequency of a quasi-circular inner binary becomes $10$Hz 
($\sim$lower end of the advanced detectors) at the critical separation
$a_1=a_{1cr}\equiv 34.6$.

Since observed NSBs have nearly equal masses (with 
relative 
difference of $\lsim 7\%$ \cite{Lorimer:2008se}),  we mainly set 
$m_0=m_1=0.1$ in  geometrical units.   For an isolated 
binary with a semimajor axis $a=0.05$AU and masses $m_0=m_1=1.4 M_\odot$, the merger time due to GW 
emission becomes  $1.0 \times 10^{10}$yr even for $e=0.7$.

As mentioned earlier, the octupole term  $H_{oc}$ vanishes for $m_0=m_1$. In order to safely
estimate its potential effects, we also examine the  case 
$(m_0,m_1)=(0.11,0.09)$.  


\subsection{Results with the Secular Theory}

As an example for predictions of the secular theory, in Fig.1, we provide the 
inner semimajor axis $a_1$ and pericenter distance $r_{p1}\equiv a_1(1-e_1)$, 
as functions of the outer cycles $N_2$. Their ratio $r_{p1}/a_1$ is identical to $(1-e_1)$.  The basic parameters for this 
calculation are given in the caption.  

\begin{figure}[]
\begin{center}
\includegraphics[width=7.cm,clip]{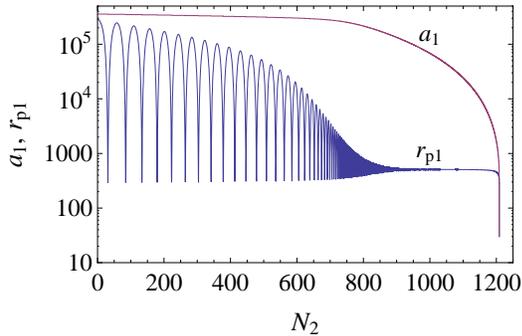}
\end{center}

\vspace*{-0.2cm}

\caption{Evolution of the inner semimajor axis $a_1$ and  pericenter distance 
$r_{p1}=(1-e_1)a_1$. 
 These results are obtained with the 
traditional secular theory.   We set $m_1=m_2=0.1$, $m_2=0.8$ with  initial 
inclination  $I_i=91^\circ$,  and initial eccentricities $e_1=0.2$ and $e_2=0.6$. 
The inner binary merges at the outer cycles  $N_{2m}=1209$.  }
\label{t1}
\end{figure}

The inner binary merges at $N_{2m}=1209$ that is considerably smaller than the 
cycles $N_{2m}=O(10^{10-11})$ for isolated binaries with moderate initial 
eccentricities \cite{Miller:2002pg,Wen:2002km}.  Due to the Kozai mechanism,  the inner eccentricity $e_1$ 
oscillates in the rangle $0.2\lsim e_1 \lsim 0.9992$, and  the minimum pericenter distances becomes $r_{p1}\simeq 300$.

When we switch off the radiation reaction and also drop the octupole and higher terms, 
we have conserved quantities in the secular theory, as mentioned after Eq.(5) (in 
particular $\sqrt{a_2(1-e_2^2)}$).  Theses conserved quantities  actually allows us to set a lower limit 
$r_{p1}\sim 300$ close to Fig.1 (see {\it e.g.} 
\cite{Wen:2002km} for the role of the 1PN effect).

In Fig.1, 
the energy of 
the inner binary is radiated mostly around the close approaches $d_1\sim 300$.   As discussed in the 
literature \cite{Miller:2002pg,Wen:2002km},  the oscillation amplitude of $e_1$ decreases gradually due to the 1PN apsidal 
precession (the last term in Eq.(\ref{om})), and, at $N_2\gsim 1000$, the inner elements 
evolve, as if an isolated binary.  The binary becomes nearly circular at the 
final phase close to the merger. 
At the 
critical separation $a_1=a_{1cr}$,  the residual eccentricity becomes $e_{1cr}=5.3 \times 10^{-3}$.

In Fig.2, using the  symbols on the solid lines, we show the duration 
$N_{2m}$ and the residual eccentricity $e_{1cr}$ at $a_1=a_{1cr}$ for  $I_i\sim 90^\circ$. The results (circles) for $(m_0,m_1)=(0.1,0.1)$ are 
similar to those (triangles) for $(m_0,m_1)=(0.11,0.09)$.  Therefore,  for the present parameters, the octupole term plays a minor role, and the 
perturbative expansion itself is effective for the secular theory (see also \cite{naoz,katz}).

\begin{figure}[h]
\begin{center}
\includegraphics[width=8.cm,clip]{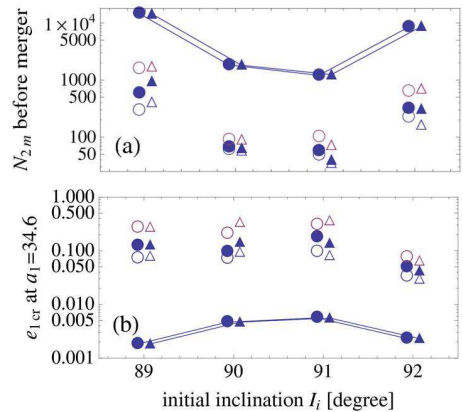}
\end{center}

\vspace*{-0.2cm}

\caption{(a) The circles ($m_0=m_1=0.1$) and triangles ($m_0=0.11,m_1=0.09$) 
represent the outer revolution cycles $N_{2m}$   before the mergers of the inner 
binaries  (slightly 
displaced  horizontally to prevent overlaps of symbols). 
 The symbols with lines 
are obtained from the traditional secular theory.  These without lines are from direct three-body calculations.   For each inclination
$I_i$, totally fifty runs with random initial mean anomalies  are 
analyzed, and we show  the median 
values (filled symbols), first (25\%) and third (75\%) quantiles (open symbols).  (b); The 
residual eccentricity $e_{1cr}$ of the inner binaries at the semimajor axes 
 $a_1=a_{1cr}=34.6$ (corresponding to the primary GW frequency of $10$Hz 
for a NSB). 
 }
\label{t2}
\end{figure}

\subsection{Direct Three-Body Calculations}

Now we move to direct three-body calculations. We use  PN equations of 
motions for spinless three-body systems, and handle the three particles 
equivalently.  In addition to the conservative terms at the  Newtonian, 1PN and 2PN orders 
(given {\it e.g.} in  \cite{Lousto:2007ji}), we included  the dissipative 2.5PN 
 terms by using Eq.(41) in \cite{Jaranowski:1996nv}.  Unless otherwise stated, we  excluded the time consuming 2PN terms 
 that would be briefly discussed later.

For numerical integration, we apply a fourth-order Runge-Kutta scheme with an 
adaptive time-step control \cite{Seto:2013an}. We terminate our runs, when the inner semimajor axis 
decreases to $a_1=a_{1cr}$ or when the instantaneous separation $d_1$ becomes 
less than $10M_1$.  The later condition reflects our perturbative (PN) 
treatment of nonlinear gravity, but no run encountered this condition. For 
numerical evaluation of the orbital elements $a_j$ and $e_j$ ($j=1,2$), we use the consecutive maximum 
($(1+e_j)a_j$) and minimum ($(1-e_j)a_j$) of the 
instantaneous orbital separations $d_j$. 

For the direct calculations, we need to specify the initial mean anomalies 
$l_j$.  
Since three-body problem depends strongly on initial conditions,  we randomly distribute the 
initial mean anomalies to examine statistical trends of evolutions. For each initial inclination $I_i$ and mass combination 
in Fig.2,  we made 50 runs, and evaluated their median values and  first/third 
quantiles of the durations $N_{2m}$ and the residual eccentricities $e_{1cr}$. 
 For $m_0=m_1=0.1$ and $I_i=90^\circ$,  we additionally 
made 50 runs,  including the 2PN terms, and obtained the median values 
$N_m=78.3$ and $e_{1cr}=0.136$ that are close to the corresponding ones in Fig.2. Therefore, for our analyses, the 2PN effect would 
not be important.

\begin{figure}[]
\begin{center}
\includegraphics[width=7.cm,clip]{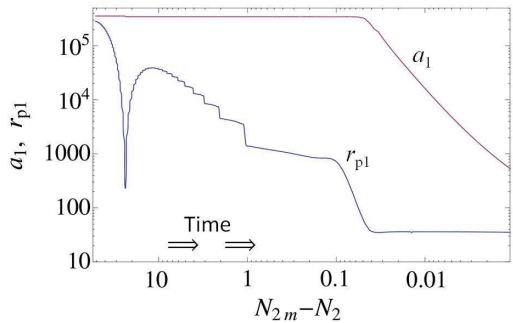}
\end{center}

\vspace*{-0.2cm}

\caption{Similar to Fig.1, but given from a direct three-body calculation. 
The inner binary takes $e_{1cr}=0.313$  at $a_1=a_{1cr}$, and merges at the outer 
cycle $N_{2m}=52.5$.  For the horizontal axis, we use the remaining outer cycle $\Delta N\equiv N_{2m}-N_2$ before the inner merger.  
}
\label{t3}
\end{figure}

We found that, in the direct calculations,  the outer parameters $a_2$ and $e_2$ 
stay nearly at their initial values, in agreement with the secular theory. 
However, Fig.2 shows that the duration $N_{2m}$ and residual $e_{1cr}$ are 
totally different.  \footnote{We have varied  the initial  parameter  $\alpha$ 
up to 1/20 (with fixing $a_2=a_{2i}$) and our main results remain true.}

To closely look at these discrepancies, we an illustrative 
 sample among the 50 runs for $I_i=91^\circ$ and $m_0=m_1=0.1$. This run ended 
 at $N_{2m}=52.5$ with the residual $e_{1cr}=0.313$  (close to the upper quantile 
 in Fig.2).
If we  simply use the outer cycle $N_2$ (as 
in Fig.1),  the semimajor axis $a_1$ comes to appear merely as a step function, and we 
cannot resolve its  rapid final evolution. Therefore, for 
Fig.3, we plot, on  a logarithm scale,  the remaining cycles $\Delta N\equiv N_{2m}-N_2$ 
before the merger. 

We can see that, up to $\Delta N=O(0.1)$,  the axis $a_1$ is nearly a 
constant, but the pericenter distance $r_{p1}$ has a modulation period  $\sim 
P_2$, the orbital period of the outer binary.  
This reflects the eccentric motion of the outer point mass $m_2$ characterized 
by $l_2$, rather than an effective ring in the secular theory. 
  Temporally neglecting radiation reaction, we follow \cite{Katz:2012je} and 
  briefly discuss  the impacts of this discreetness for 
evolution of the inner specific angular momentum vector $\vej_1$ 
(closely related to $e_1$ and $r_{p1}$ as $|\vej_1|=\sqrt{a_1(1-e_1^2)}$).   Its
variation $\Delta \vej_1$ due to $m_2$ in  one inner orbital 
revolution, 
depends strongly on the exact position of $m_2$ and thus has a 
stochastic character (denoting its rms value by $\delta j_1$).

For $\delta j_1\ll |\vej_1|$, the total variation of $\vej_1$ after a few outer 
orbital cycles 
could be close to that caused by the corresponding 
 outer ring, and the orbital averaging  could be efficient. However, 
for a highly eccentric case with  $\delta j_1\gsim |\vej_1|$, the averaging method would 
break down, and consequently, the associated lower limit for  
$r_{p1}$ (mentioned in \S III.A) would 
be no longer valid.  In the direct three-body integral, the discreetness of $m_2$ 
is naturally included, and we have possibilities to realize $r_{p1}$ smaller  
than the limit obtained with the secular theory. While we temporally neglected 
radiation reaction for simplicity, we can expect similar differences for our 
dissipative systems.

Indeed, in Fig.3, at the turning point $\Delta N\sim 4\times 10^{-2}$,  the quantity $1-e_1$ takes a minimum value, 
corresponding to $r_{p1}=38$ (much smaller than Fig.1).  Then   
 the inner binary evolves almost 
independently of the outer body $m_2$ with rapidly decreasing $a_1$ from 
$a_1=3.0\times 10^5$ but nearly conserving $r_{p1}$ for a while.

For an orbit with $1-e_1\ll 1$,  GW emission is 
dominated at the pericenter passages, and the radiated energy there is given as
$
\delta E\sim -{85\pi}{(m_0 m_1)^2 M_1^{1/2}}/(12\sqrt{2}{r_{p1}^{7/2}})$, depending 
  strongly on $r_{p1}$ \cite{O'Leary:2008xt}.
At the turning point in Fig.3, this amounts to a fraction
\beq
Y\sim 0.19 \lmk\frac{a_1}{3.0\times 10^5}\rmk \lmk\frac{r_{p1}}{38}\rmk^{-7/2}\label{rgw}
\eeq
  of the inner orbital energy $-m_0 m_1/2a_1$.

Meanwhile, for the secular theory,  we can simply estimate the local minimum of $1-e_1$ 
from Eq.(\ref{de}) (with $de_1/dt=0$) \cite{Wen:2002km}.
This is determined by the balance between the two effects, the dissipative radiation 
reaction working only around $d_1=O(r_{p1})\ll a_1$ and the 
tidal effect (by $m_2$) operating mainly during $d_1=O(a_1)$.  
Neglecting the octupole terms,  we 
obtain the minimum pericenter distance $r_{p1,min}=a_1 (1-e_1)_{min}$ as
\beq
r_{p1,min}\simeq 60 \lmk\frac{a_2/a_1}{71}\rmk \lmk\frac{a_1}{3.0\times 10^5}\rmk^{1/6}
\lmk \frac{X}1\rmk^{-1/3}\label{peri2}
 \eeq
with  the factor
$
X\equiv \sin^2 I |\sin 2\omega_1 | \le 1
$.  
Thus, even with the highly conservative setting $X=1$, the distance $r_{p1}= 38$ at 
the turning  point in Fig.3 is not allowed in the secular theory, and the 
radiated fraction becomes at most $Y=0.02$, in contrast to 
Eq.(8). Eq.(9) has been used in previous studies, with additionally evaluating  
$X$ \cite{Wen:2002km}.
But, along with the insufficient treatment of the discreteness effect (mentioned earlier), 
the two temporally separated effects are directly compared in Eq.(9) without resolving the inner 
orbital phase. Roughly speaking, even at $d_1\gg r_{p1}$, a nearly radial inner 
orbit could be prohibited by the radiation reaction that intrinsically has no effect there.  


\section{Discussions}

Finally, we comment on the implications  of our results for GW astronomy.
In Fig.3, after the turning point,     the inner binary emits pulse-like GWs 
around the pericenter passages \cite{Gould:2010ij}. This waveform 
has a characteristic frequency $({M_1/r_{p1}^3})^{1/2}/\pi\sim 10$Hz that  is $\sim 30$ times higher than the counterpart in 
Fig.1.  While Figs.1 and 3 are given for a  specific set of 
parameters, this shift would be encouraging for ground-based GW observation, given 
the formidable noise walls  below $\sim 10$Hz \cite{adv}.

Fig.2 shows that we could have  larger residual eccentricities $e_{1cr}$ and also 
shorter merger times  than the estimations by the  secular theory. These differences are 
closely related to the decrease of the pericenter distances, and suggest 
a higher merger rate of NSBs in star clusters, as discussed earlier.
For a quasi-circular binary,  the residual eccentricity could be probed through 
the associated phase modulation of inspiral GWs \cite{Krolak:1995md}.  For a NSB 
detectable with advanced detectors at  SNR$\sim 15$,  the 
 resolution of the residual value $e_{1cr}$  (at $10$Hz) would be $\Delta e_{1cr}\simeq 
0.01$ \cite{Krolak:1995md}.  Interestingly, this is just between the two 
predictions in Fig.2 and  we might  discriminate  the origins of NSB 
mergers with the upcoming GW detectors.

\acknowledgements
This work was supported by JSPS (20740151, 24540269) and
MEXT (24103006).


\begin{thebibliography}{}


\bibitem{adv}
K. Kuroda: LCGT collaboration, Class. Quantum Grav.
27, 084004 (2010); Advanced LIGO 
http://www.cascina.virgo.infn.it/advirgo/; Advanced Virgo http://www.cascina.virgo.infn.it/advirgo/.



\bibitem{Lorimer:2008se} 
  D.~R.~Lorimer,
  Living Rev.\ Rel.\  {\bf 11}, 8 (2008).


\bibitem{Krolak:1995md} 
  A.~Krolak, K.~D.~Kokkotas and G.~Schaefer,
  Phys.\ Rev.\ D {\bf 52}, 2089 (1995).

\bibitem{Blaes:2002cs} 
  O.~Blaes, M.~H.~Lee and A.~Socrates,
  Astrophys.\ J.\  {\bf 578}, 775 (2002).


\bibitem{Miller:2002pg} 
  M.~C.~Miller and D.~P.~Hamilton,
    Astrophys.\ J.\  {\bf 576}, 894 (2002).

\bibitem{Wen:2002km} 
  L.~Wen,
  Astrophys.\ J.\  {\bf 598}, 419 (2003);
  T.~A.~Thompson,
  Astrophys.\ J.\  {\bf 741}, 82 (2011).

\bibitem{Kozai:1962zz} 
  Y.~Kozai,
  Astron.\ J.\  {\bf 67}, 591 (1962); M.  L. Lidov, Planet.Space Sci., {\bf 
	9},719 (1962).


\bibitem{gs}
H.  Goldstein, {\it Classical Mechanics, second edition}  (Addison-Wesley, 1980).


\bibitem{Ford:1999mi} 
  E.~B.~Ford, B.~Kozinsky and F.~A.~Rasio,
  Astrophys.\ J.\  {\bf 605}, 966 (2004).


\bibitem{Antonini:2012ad} 
  F.~Antonini and H.~B.~Perets,
  Astrophys.\ J.\  {\bf 757}, 27 (2012).



\bibitem{ssd}
C. D. Murray and  S. F.  Dermott, {\it Solar System Dynamics}  (Cambridge University Press, UK, 1999).


\bibitem{Katz:2012je} 
  B.~Katz and S.~Dong,
  arXiv:1211.4584.



\bibitem{holman} 
  M.~Holman, J. Touma and S. Tremaine, Nature {\bf 386}, 254 (1997).

\bibitem{naoz} 
  S.~Naoz et al., Nature {\bf 473}, 187 (2011).



\bibitem{Peters:1964zz} 
  P.~C.~Peters,
  Phys.\ Rev.\  {\bf 136}, B1224 (1964).



\bibitem{katz} 
  B.~Katz, S.~Dong and R.~Malhotra,
  Phys.\ Rev.\ Lett.\  {\bf 107}, 181101 (2011).

\bibitem{Lousto:2007ji} 
  C.~O.~Lousto and H.~Nakano,
  Class.\ Quant.\ Grav.\  {\bf 25}, 195019 (2008).

\bibitem{Jaranowski:1996nv} 
  P.~Jaranowski and G.~Schaefer,
  Phys.\ Rev.\ D {\bf 55}, 4712 (1997).



\bibitem{Seto:2013an}
N.~Seto,
  Phys.\ Rev.\ D {\bf 85}, 064037 (2012);
  N.~Seto,
  arXiv:1301.3135.

\bibitem{Gould:2010ij} 
  A.~Gould,
  arXiv:1011.4518 [astro-ph.SR].

\bibitem{O'Leary:2008xt} 
  R.~M.~O'Leary, B.~Kocsis and A.~Loeb,
  Mon.Not.Roy.Astron.Soc.   {\bf 395}, 2127 (2009);
B.~Kocsis and J.~Levin,
  Phys.\ Rev.\ D {\bf 85}, 123005 (2012).





\end{thebibliography}
\end{document}